\documentclass[lettersize,journal]{IEEEtran}
\usepackage{amsmath,amsfonts}
\usepackage{algorithmic}
\usepackage{algorithm}
\usepackage{array}
\usepackage{textcomp}
\usepackage{stfloats}
\usepackage{url}
\usepackage{verbatim}
\usepackage{graphicx}
\usepackage{cite}
\usepackage[nolist]{acronym}
\usepackage{float}
\usepackage{makecell}
\usepackage{subcaption}

\begin{document}

\title{Dynamic Bayesian Network Modelling of User Affect and Perceptions of a Teleoperated Robot Coach during Longitudinal Mindfulness Training}





\author{Indu P. Bodala$^{1}$ and Hatice Gunes$^{2}$
\vspace{-5mm}
\thanks{$^{1}$Department of Electronics and Computer Science, University of Southampton, United Kingdom. {\tt\small\ i.p.bodala@soton.ac.uk}}%
\thanks{$^{2}$Department of Computer Science and Technology, University of Cambridge, United Kingdom. {\tt\small\ hatice.gunes@cl.cam.ac.uk}}%
\thanks{This work has been submitted to the IEEE for possible publication. Copyright may be transferred without notice, after which this version may no longer be accessible.}
}

\maketitle

\vspace{-5mm}
\begin{abstract}
Longitudinal interaction studies with Socially Assistive Robots (SARs) are crucial to ensure that the robot is relevant for long-term use and its perceptions are not prone to the novelty effect. In this paper, we present a dynamic Bayesian network (DBN) model to capture the longitudinal interactions participants had with a teleoperated robot coach (RC) delivering mindfulness sessions. DBN modeling is used to study complex, temporal interactions between the participants’ self-reported personality traits, weekly baseline wellbeing scores and session ratings, and participants' facial AUs elicited during the sessions during a 5-week longitudinal study. The learnt model is an intuitive graphical representation that captures the following within- and between-sessions aspects of the longitudinal interaction study – $(i)$ influence of the 5 personality dimensions on the facial AU states and the session ratings; $(ii)$ influence of facial AU states on the session ratings; and $(iii)$ influences within the items of the session ratings. The DBN model was learnt using the first $3$ weeks and is used to predict the session ratings of the remaining $2$ weeks of the $5$-weeks longitudinal data. The goodness-of-fit of the model is quantified in terms of subject-wise $RMSE$ and $R^2$ scores. We demonstrate two applications of the model – imputation of missing values and estimation of longitudinal session ratings of a new participant with a given personality profile. The DBN model thus facilitates learning of conditional dependency structure between variables in the longitudinal data and offers inferences and conceptual understanding which is not possible through other regression methodologies.
\end{abstract}

\begin{IEEEkeywords}
Dynamic Bayesian Networks, User Affect, User Perceptions, Longitudinal HRI, Mindfulness, Teleoperation.
\end{IEEEkeywords}

\section{Introduction}
\acfp{SAR} have the potential of becoming companions that can assist in maintaining wellbeing and provide timely advice for the upkeep of mental health \cite{feil2005defining}. Moreover, the embodied multi-modal interactions based on speech, facial affect and gestures exhibited by the robot companions can better engage users physically and socially compared to digital interfaces such as mobile phones and computers \cite{conti2021brief, lima2021robotic}. Recent \acf{HRI} studies have used robots for delivering mental wellbeing interventions such as positive psychology and mindfulness \cite{jeong2020robotic, bodala2021teleoperated}. Despite achieving positive results in assistive interventions, \acsp{SAR} face a major challenge in terms of their acceptance for long-term use \cite{gockley2005designing, de2017they}. Moreover, current studies in the field of \ac{HRI} are undertaken as one-off or short-term interactions, and longitudinal studies are sparse. There is also a need to deploy appropriate methodology to analyze the interaction data obtained from longitudinal studies. 

Longitudinal investigation of interactions with social robots is crucial in understanding how the users perceive the robot functionality beyond the initial impressions and experiences also known as the `novelty effect' \cite{paetzel2020persistence}. Researchers have shown that when the novelty effect wears off, people lose interest and change their attitudes towards the robots resulting in a decreased usage \cite{winkle2018social, de2017they}. Recent works on longitudinal investigation also offer insights beyond the issue of novelty effect. \cite{bodala2021teleoperated} studied how interacting with a teleoperated robot across time leads to improved perception ratings of the robot functionality such as robot motion and conversation. Longitudinal investigations could also provided strategies for betterment of human-robot teaming \cite{de2020towards}. Additionally, social robots delivering wellbeing and rehabilitation interventions warrant an investigation into factors that contribute to adherence and behavioural change in users, which would not be possible without studying long-term interactions \cite{cespedes2021socially}. In this paper, we propose to develop a DBN model to study the evolving user affect and perceptions towards a \ac{RC} delivering weekly mindfulness sessions for $5$ weeks as a function of participants' personality, baseline wellbeing and facial AU states. 

\acfp{BN} are probabilistic graph networks that encode conditional relationships between multiple stochastic processes \cite{scutari2009learning}. They are the directed acyclic graphs where \textit{nodes} represent random variables and \textit{edges} represent conditional dependencies between them. \acp{DBN} are an extension of the static \acp{BN} that can include temporal dependencies between variables and are a useful methodology to study temporal data \cite{murphy2002dynamic}. Graphical models, specifically \acp{DBN} are highly useful for dealing with uncertainty in the temporal data, for modeling complex relationships such as multi-modal and context-driven dependencies between variables and for representation of latent processes such as mental states or affective states. These properties make \acp{DBN} suitable for studying multi-modal, temporal interaction data \cite{huang2014learning}. 

In this paper, we use \ac{DBN} modelling to analyze the data gathered from the participants during longitudinal mindfulness sessions with the \ac{RC}. We aim to achieve the following objectives from this modelling:
\begin{enumerate}
    \item to develop an intuitive probabilistic model that can encode the longitudinal changes in the participants' session ratings comprising perceptions of the RC and the self-reported change in relaxation/calm levels due to the mindfulness training, and the influence of participants' personality, baseline wellbeing measures and their facial AU states on these session ratings;
    \item to use the developed model to infer the session ratings for future time steps; and
    \item to demonstrate applications of the developed model such as imputation of missing data in the dataset and inference of longitudinal trends in the ratings of new participants with varying personality profiles. 
\end{enumerate}

\section{Related Work}

\subsection{Longitudinal Studies and Analyses}
Longitudinal studies provide insights about the impact of the robot design and interactions which are otherwise restricted by factors such as novelty and adaptation to technology \cite{leite2013social}. However, very few studies in \ac{HRI} and social robotics investigated the use of robots for more than a one-off interaction session \cite{cespedes2021socially, perugia2021can}. \cite{paetzel2019let} gradually exposed participants to the capabilities of a robot and their perception of it was tracked from their first impression to after playing a short interactive game with it. They found that the initial uncanny feelings towards the robot were significantly decreased with further interactions.

It is also crucial to select appropriate methodology to analyse longitudinal data characterized by intra- and inter-individual variability across time. Multilevel modelling methodology is effective in understanding both inter- and intra-individual variability, can handle violations in error assumptions for univariate analysis and can also incorporate multiple change predictors that are either static or dynamic, for multivariate analysis \cite{ployhart2011quick}. \cite{bodala2021teleoperated} used growth modelling to study the longitudinal changes in participants' ratings of human and teleoperated robot coaches delivering mindfulness training. \cite{spangler2020multilevel} used multilevel modelling to analyze changes in performance as a function of stress and cardiovascular response at multiple time scales such as task, block and session. \cite{mcduff2019longitudinal} used hierarchical regression to select best model from incremental linear regression models to understand multiple factors affecting longitudinal emotional progression across several weeks. These methods provide an understanding of factors affecting change in the variables of interest when viewed separately. However, to gain a holistic understanding of the causal processes underlying multiple variables of interest, we need to use methods that can encapsulate a conditional dependency structure between variables, as our observations provide information about multiple variables in the state space.

\subsection{Bayesian network modelling in affective computing and HRI}

Probabilistic models such as \acsp{DBN} have the capability to intuitively represent relationships across multiple variables and model uncertainties in these relationships. This kind of reasoning is not available using deep learning methods as they are unable reason about causal attributions, explanations or incorporate contextual knowledge into their inferences \cite{ong2019applying}. \acsp{DBN} have been used for context-aware models in affective computing studies. \cite{ong2015affective} performed complex inference of emotions of participants while gambling given the gambling outcome and other verbal cues. \cite{wu2018rational} modelled how mental states relate to emotional expressions and how they lead to actions using Bayesian inference. \cite{otsuka2007automatic} proposed a 3-layer DBN framework to evaluate the interaction between regimes and behaviors during multi-party conversations where the first layer perceives speech and head gestures, the second layer estimates gaze patterns while the third one estimates conversation regimes. \cite{mihoub2016graphical} developed a multimodal behavioral model based on human-human interactions using \ac{DBN} to generate coverbal actions (gaze, hand gestures) for the subject given verbal productions, the current phase of the interaction and the perceived actions of the partner. 

Context-aware modelling using \acsp{BN} was also deployed in \ac{HRI} studies to infer robot and/or user actions as well as teaming strategies. \cite{hong2007mixed} modelled a hierarchical Bayesian networks manually for a mixed-initiative interaction of human and service robot. \cite{montesano2008learning} used \acsp{BN} for the task of learning object affordances by social robots where the affordances are encoded in the probabilistic relations between actions and perceptions (object features and effects). \cite{song2015task} presented a probabilistic framework for the representation and modelling of robot-grasping tasks. This facilitated inference of grasps from several different combinations of observations where the estimation task was expected to be highly multi-modal, i.e., that there are many possible grasps that afford a specific task. In this study, we use DBN modelling to understand longitudinal changes of participants' perceptions towards a teleoperated robot mindfulness coach across multiple interactions as a result of various factors such as participants' personalities, baseline wellbeing and facial AU states displayed during the interactions.

\section{Materials and Methods}

In this section, we present the details of the mindfulness study involving longitudinal interactions with a \acf{RC}, data collection and the \ac{DBN} methodology for the data analysis.

\subsection{Experimental design}
\begin{figure}
  \centering 
  \includegraphics[width=\columnwidth]{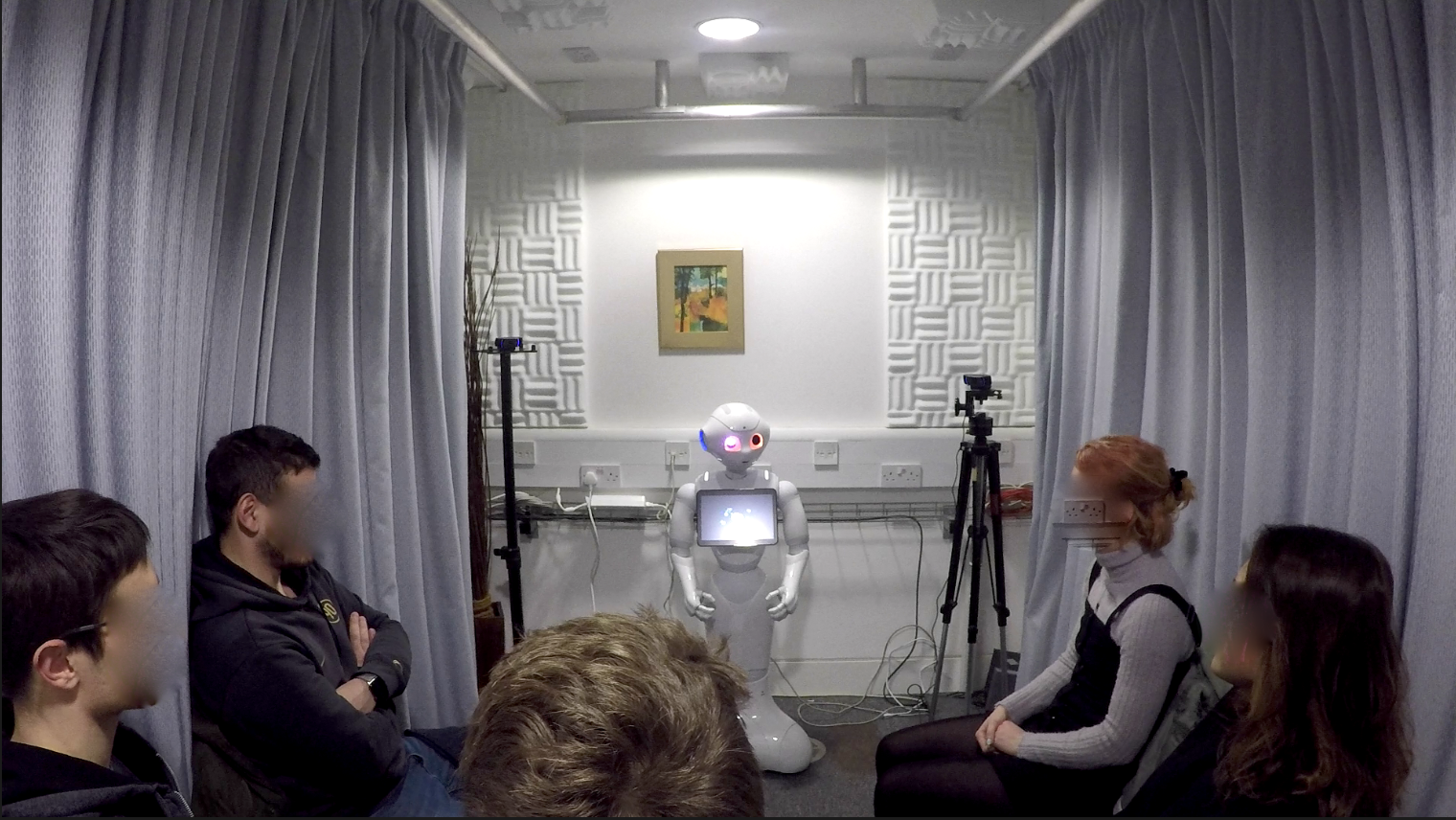}
  \caption{Participants interacting with the teleoperated mindfulness robot coach.}
  \label{fig:exp_set_a}
\end{figure}

\begin{figure}
  \centering 
  \includegraphics[width=\columnwidth]{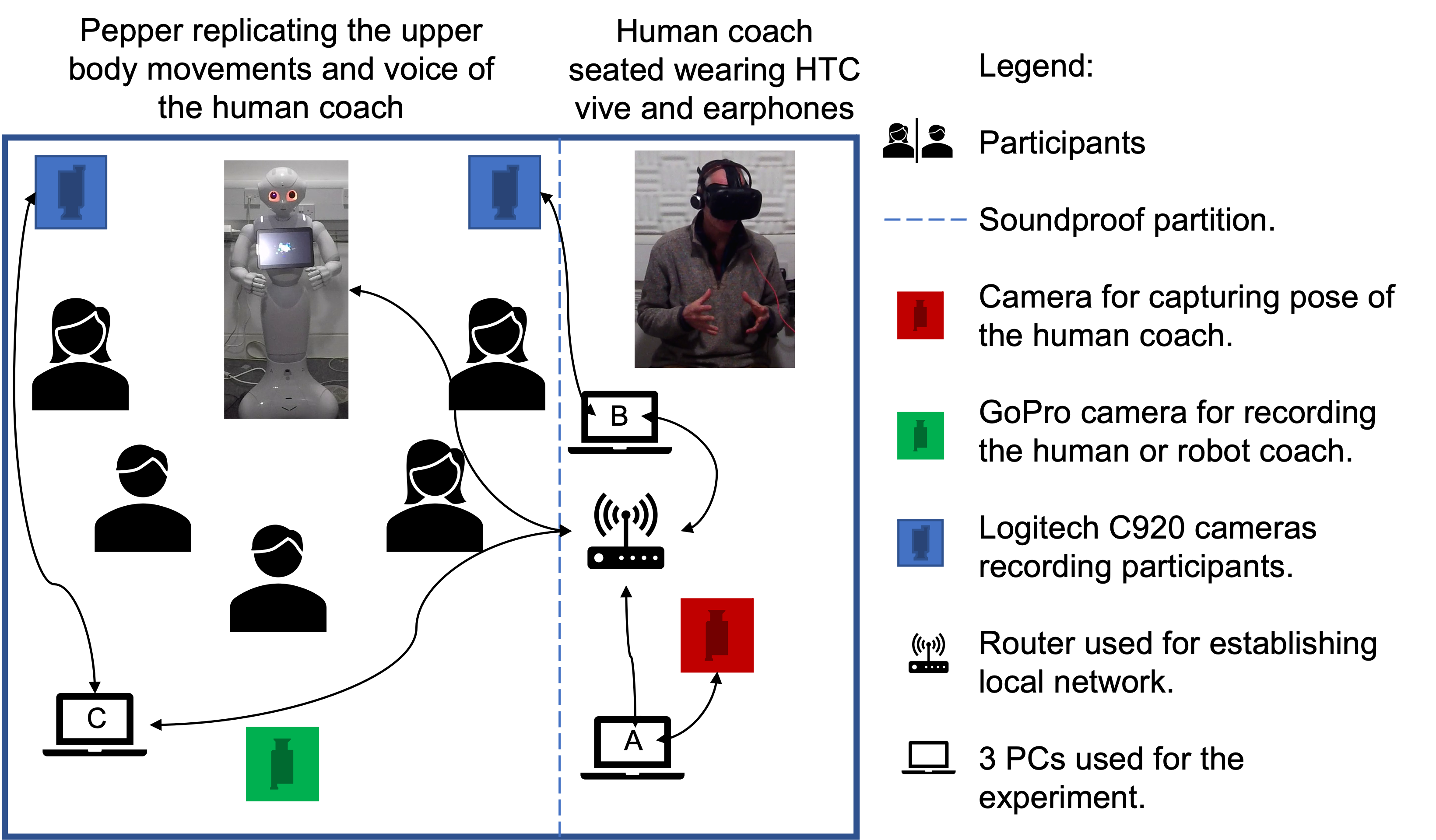}
  \caption{Teleoperation setup, where an experienced human coach is remotely teleoperating Pepper through real-time pose replication. The HMD worn by the human coach enables them to see what the robot sees in the adjacent room. An audio pipeline is also created to enable the coach to have a conversation with the participants.}
  \label{fig:exp_set_b}
\end{figure}

The study was approved by the Ethics Committee of the Department of Computer Science and Technology, University of Cambridge. All participants provided informed consent for data collection after reading the provided Information Sheet describing the study details. Participants were advised that they should not be undertaking other professional mental health related treatments or medication to take part in this study. We also asked the participants to fill the Participant Health Questionnaire (PHQ9)~\cite{kroenke2001phq} and the General Anxiety Disorder (GAD7)~\cite{spitzer2006brief} questionnaire to assess and screen for high depression and anxiety levels as our target group was a non-clinical population. We conducted a 5-week longitudinal study where two groups of participants, $4$ in one one group and $5$ in the other, with a total of $9$ participants ($3$ females)\footnote{\scriptsize{Our initial plan to recruit a higher number of participants was interrupted by the COVID-19 pandemic.}} received mindfulness training from the RC as shown in Fig. \ref{fig:exp_set_a}. The teleoperation platform enabled an experienced human coach to deliver mindfulness sessions remotely through the Pepper robot\footnote{\scriptsize{https://www.softbankrobotics.com/emea/en/pepper}} with the help of 3 pipelines -- a pose replication pipeline where human coach's pose is replicated onto the robot in real-time, a vision pipeline where the images captured by the camera on the robot's forehead were projected onto a head-mounted display worn by the human coach so that they can see what the robot was seeing, and an audio pipeline that enabled the human coach to converse with the participants from a remote location. $5$ out of $9$ participants attended all $5$ sessions while others missed one session each, at random. Each weekly session was administered for $\approx40$ minutes to each group where the robot was teleoperated by the human coach Fig. \ref{fig:exp_set_b}. The course structure was based on the 'Mindful Student Study' \cite{galante2018mindfulness}. The group sessions did not just contain meditations or breathing exercises but also included group interactions between the RC and the participants, including discussions on the importance of these practices, how the participants felt about them, and whether and how they could be integrated into their daily life. Further details of the developed teleoperation platform and the session structure were provided in \cite{bodala2021teleoperated}. Based on the session structure, the experimenter coded the length of each session into the following sub-sessions: \textit{Introduction}, where the coach welcomed the participants and introduced the topic for the week; \textit{Meditation}, the period during which the participants did meditation practice with eyes closed; \textit{Activity}, the period during which the participants engaged in activities such as writing about thoughts, charting the focus levels, paying attention while eating a chocolate, etc.; \textit{Interaction}, during which the participants interacted with the coach, shared thoughts about their experience, and asked questions about the meditation practice; and \textit{Conclusion}, where the coach sums up the discussion and advises participants on how they can carry out the practice at home. All weekly sessions contained these sub-sessions except the \textit{Activity}, which is present in only some of the sessions.

\vspace{-2mm}

\subsection{Data}

\paragraph{Questionnaires} Each participant filled a $20$-item personality questionnaire \cite{topolewska2014short} on the scales of \textit{Extroversion, Agreeableness, Conscientiousness, Neuroticism and Openness}. At the beginning of each weekly session, they also filled \acf{WEMWBS} questionnaire \cite{tennant2007warwick} to reflect the baseline wellbeing level at that time. After each session, the participants filled a `session experience questionnaire' evaluating their experience interacting with the \ac{RC}. The questionnaire was adapted from a combination of the Godspeed \cite{bartneck2009measurement} and the human-robot interaction questionnaires \cite{romero2015testing}. While the former evaluates the participants' impressions of a robot in terms of \textit{Anthropomorphism, Animacy, Likeability} and \textit{Perceived Intelligence}, the latter measures the interactions with items based on \textit{Robot Motion, Conversation} and \textit{Sensations}. Additionally, we also measured how the participants felt at the beginning and end of each session on the scales: \textit{Anxious -- Relaxed} and \textit{Agitated -- Calm}, and the difference between these beginning and the end ratings is denoted as $\Delta RC$

\begin{figure*}[ht]
  \centering
  \includegraphics[width=\textwidth]{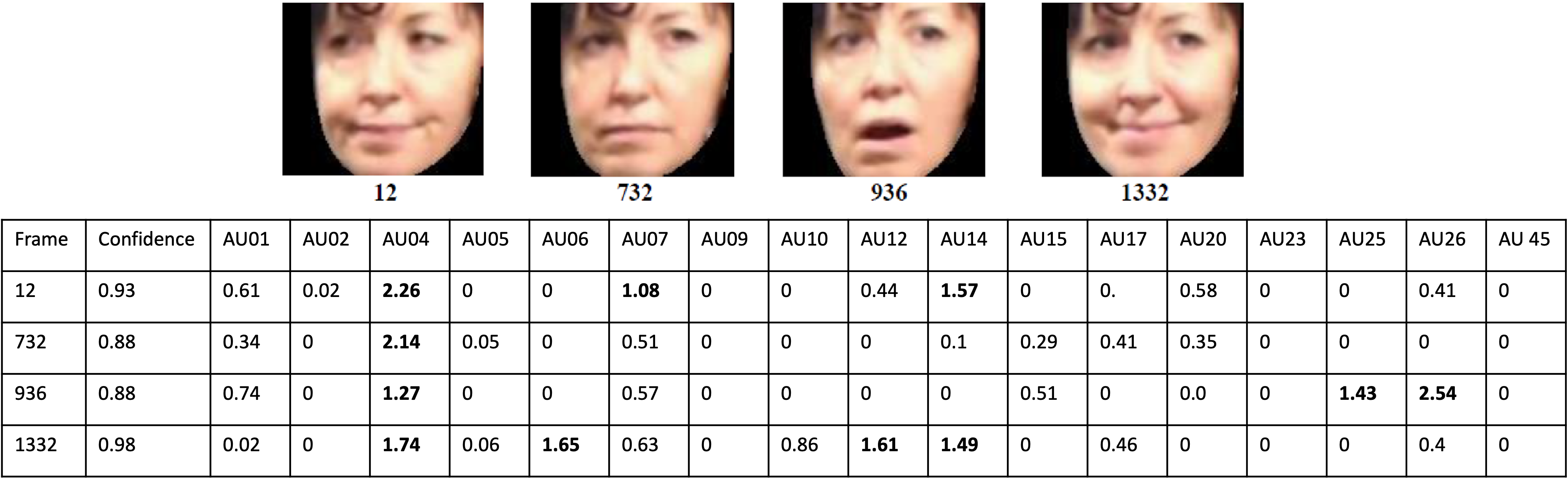}
  \caption{AU intensities corresponding to the 18 AUs extracted for a selection of $4$ frames are shown here. The intensities ranged from $0$ (not present), $1$ (present at minimum intensity), $5$ (present at maximum intensity), with continuous values in between. We only considered AU intensities greater than $1$ as it corresponds the minimum intensity value, if the AU is present. AUs active for Frame $12$ are Brow Lowerer (AU04), Lid Tightener (AU07) and Dimpler (AU14); for Frame $732$ are Brow Lowerer (AU04); for Frame $936$ are Brow Lowerer (AU04), Lips part (AU25) and Jaw Drop (AU26); for Frame $1332$ are Brow Lowerer (AU04), Cheek Raiser (AU06), Lip Corner Puller (AU12) and Dimpler (A14).}
  \label{fig:AU_example}
  \vspace{-2mm}
\end{figure*}

\paragraph{Facial Data} We used the video recordings of each cropped individual participant's face to obtain \acf{AUs} with the help of OpenFace toolbox \cite{baltruvsaitis2016openface}. Intensity values of 18 \ac{AUs}\footnote{\scriptsize{List of 18 AUs used: https://github.com/TadasBaltrusaitis/OpenFace/wiki/Action-Units}} for each participant were obtained per each frame of the video recording. Fig. \ref{fig:AU_example} demonstrates facial AUs corresponding to $4$ frames from a participant. We extracted $4$ summary measures, mean, max, min and standard deviation of the intensity values for each of the 18 \ac{AUs} for each sub-session, i.e. the facial features from a participant is an array of length $18\times4=72$ elements for each sub-session.

\begin{figure}
  \centering 
  \includegraphics[width=0.9\columnwidth]{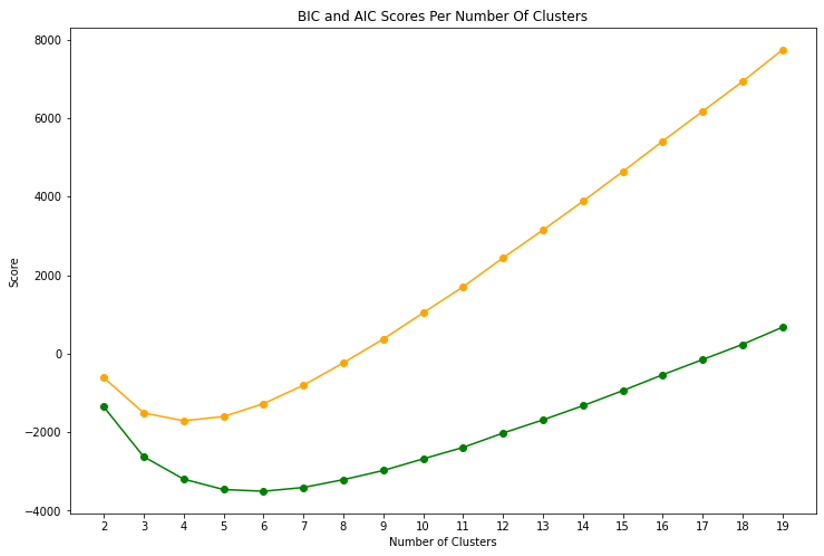}
  \caption{Plot of BIC (in \textit{orange}) and AIC (in \textit{green}) scores against  number of clusters $AU\_med$. BIC scores were found to be minimum for the cluster number $K=4$.}
  \label{fig:bic_a}
\end{figure}

\begin{figure}
  \centering 
  \includegraphics[width=0.9\columnwidth]{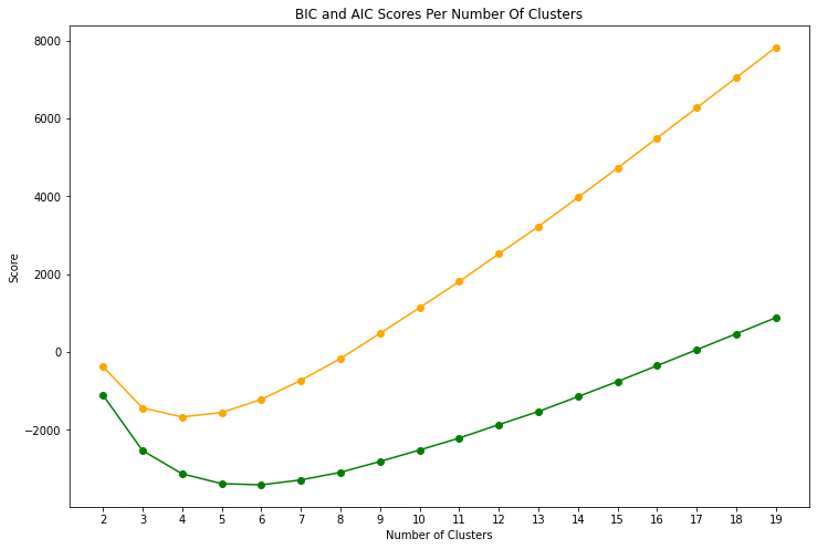}
  \caption{Plot of BIC (in \textit{orange}) and AIC (in \textit{green}) scores against  number of clusters for $AU\_int$. BIC scores were found to be minimum for the cluster number $K=4$.}
  \label{fig:bic_b}
\end{figure}

For the \ac{DBN} model, it was desirable to represent the facial data using lesser number of nodes. Hence, we performed Gaussian mixture model (GMM) clustering\footnote{\scriptsize{https://scikit-learn.org/stable/modules/mixture.html}}, a probabilistic soft clustering approach for distributing the points in different clusters, to discretize the values for the nodes corresponding to the facial data for the \textit{Meditation} and the \textit{Interaction} sub-sessions separately. The optimum number of clusters, $K$, was decided as the number that minimizes the BIC scores. The AIC and BIC scores plotted for a range of $2-20$ clusters. For both the \textit{Meditation} (Fig. \ref{fig:bic_a}) and the \textit{Interaction} (Fig. \ref{fig:bic_b}) sub-sessions, the BIC score was found to be minimum at $K=4$ clusters. Hence, the nodes representing facial features of the participants during \textit{Meditation} and the \textit{Interaction} sub-sessions were denoted as $AU\_med$ and $AU\_int$ respectively and could take 4 values $(1, 2, 3$ and $4)$ depending on which cluster the feature array belonged to.

\vspace{-3mm}

\subsection{\acf{DBN} Learning and Inference}

To facilitate learning of graphical structures for longitudinal data, we defined a 2 time-slice \ac{DBN}. A 2 time-slice structure implies that the nodes at the time $t$ are influenced by the values of the nodes at the time-points $t$ and $t-1$ only. We used \textit{bnstruct} package \cite{franzin2017bnstruct} in R to perform learning and inference on the \ac{DBN}. 

\begin{figure*}[ht]
  \centering
  \includegraphics[width=\textwidth]{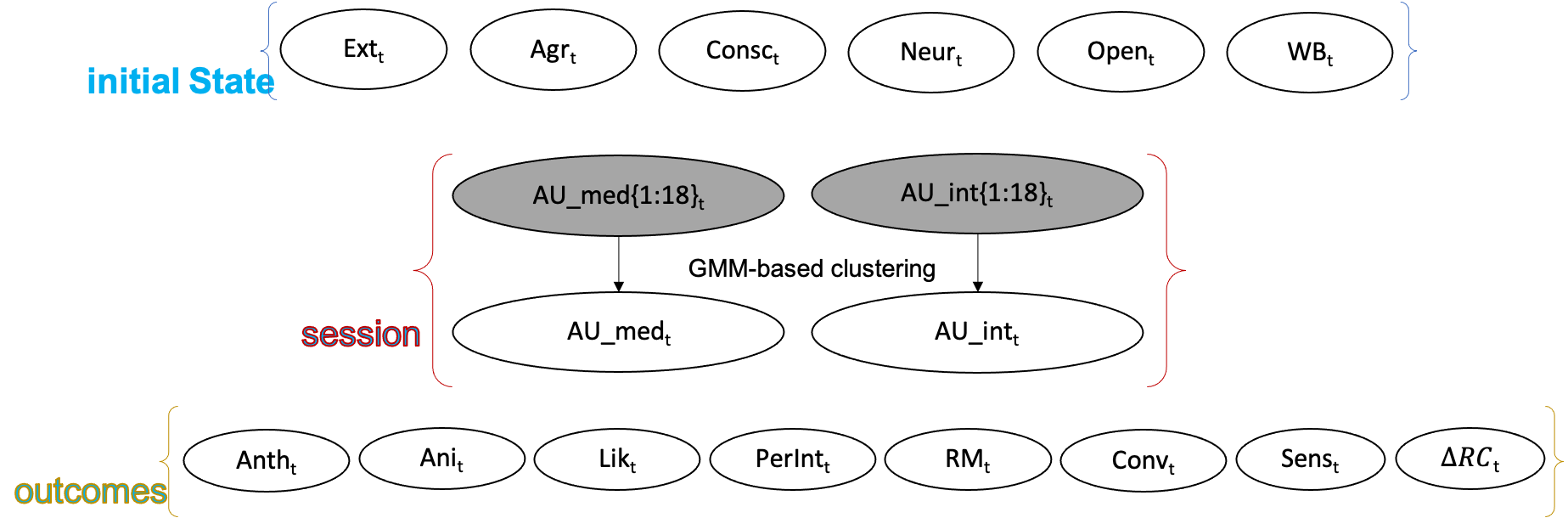}
  \caption{Nodes of the \ac{DBN} for $1$ time-slice, representing data acquired from $1$ mindfulness session. The nodes are further divided into 3 layers -- first layer comprises the \textbf{initial state} of the participant given by their personality scores and baseline wellbeing level; second layer comprises the facial AU states experienced by the participants during the mindfulness \textbf{sessions}; and the third layer comprises session ratings which are the \textbf{outcomes} of the session with the robot coach. The nodes in gray are the individual AU values that are used for clustering and not for the subsequent stages of the DBN learning.}
  \label{fig:Nodes}
\end{figure*}

\paragraph{Defining Nodes}  The nodes of the network are defined from the random variables corresponding to the data gathered in the form of questionnaires and facial AU data. Table. \ref{table:nodes} provides information of each node and Fig. \ref{fig:Nodes} shows how the nodes are arranged at time $t$. The model comprises of $16$ nodes at each time-slice $t$, which represent one week's data. 

\begin{table*}[htb]
\centering
  \caption{Details of the defined nodes of the \ac{DBN} for $1$ time-slice, i.e. for $1$ week.}
  {
  \small
  \begin{tabular}{llllll}\Xhline{1.0 pt}
    \makecell[l]{\textit{\textbf{Node Type}}} &
    \makecell[l]{\textit{\textbf{Node Name}}} & \makecell[l]{\textit{\textbf{Acronym}}} & \makecell[l]{\textit{\textbf{Remarks}}} &
    \makecell[l]{\textit{\textbf{Discreteness}}}&
    \makecell[l]{\textit{\textbf{Node Size}}}\\ \Xhline{1.0 pt}
    \makecell[l]{Personality} &
     \makecell[l]{Extroversion\\ Agreeableness\\ Conscientiousness\\ Neuroticism\\ Openness} &
    \makecell[l]{$Ext$\\ $Agr$\\ $Consc$\\ $Neur$\\ $Open$} &  \makecell[l]{Big-five personality scores.\\ Collected only once for each\\ participant in the study. \\The values are replicated\\ 
    for all 5 weeks.} &
    \makecell[l]{Continuous} &
    \makecell[l]{3}\\
    \Xhline{0.5 pt}
    \makecell[l]{Baseline Wellbeing} &
    \makecell[l]{WEMWBS score} &
    \makecell[l]{$WB$} &  \makecell[l]{Average score form WEMWBS\\ survey. \\Obtained at the beginning\\ of each session.} &
    \makecell[l]{Continuous} &
    \makecell[l]{3}\\
    \Xhline{0.5 pt}
    \makecell[l]{Facial AUs} &
    \makecell[l]{AU state during meditations\\ AU state during interactions} &
    \makecell[l]{$AU\_med$ \\ $AU\_int$} &  \makecell[l]{AU state is obtained from the\\ GMM clustering of the\\ summary measures of 18 AUs\\ of the participants.}&
    \makecell[l]{Discrete} &
    \makecell[l]{4}\\
    \Xhline{0.5 pt}
    \makecell[l]{Session experience \\ratings} &
    \makecell[l]{Anthropomorphism\\ Animacy\\ Likeability\\ Perceived Intelligence\\ Robot Motion\\ Conversation\\ Sensations\\
    } &
    \makecell[l]{$Anth$\\ $Ani$\\$Like$\\ $PerInt$\\ $RM$\\ $Conv$\\ $Sens$} &  
    \makecell[l]{Perception Ratings\\ of \ac{RC}}&
    \makecell[l]{Continuous} &
    \makecell[l]{3}\\
    \Xhline{0.5 pt}
    \makecell[l]{Session experience \\ratings}&
    \makecell[l]{Changes in \\relaxation/calm ratings} &
    \makecell[l]{$\Delta RC$} &
    \makecell[l]{Difference between \textit{Relaxation}\\ and \textit{Calm} ratings before\\ and after each session.} &
    \makecell[l]{Continuous} &
    \makecell[l]{3}\\
    \Xhline{1.0 pt}
  \end{tabular}
  }
\label{table:nodes}  
\end{table*}

\paragraph{Imputation and Node Sizes}
Our dataset contained missing values caused by participants missing a weekly session. To deal with the missing values, we used k-nearest neighbour (knn) imputation, to replace the missing values in the datasets with the mean value from the $k$ nearest neighbors found in the training set. The value of $k$ was chosen as $2$, i.e. an average of previous and next sessions.

The learning in \ac{DBN} is implemented with discrete variables alone, i.e. continuous variables are quantized after imputation. The column, \textit{Discreteness} in Table. \ref{table:nodes} states whether the input given to the model is discrete or continuous and the column, \textit{Node size} signifies the possible number of values of the node, if the node is discrete or the number of levels it may be quantized to in the later steps, if the node is continuous. The node sizes for the $AU\_med$ and $AU\_int$ were provided to be $4$, signifying the 4 discrete values these nodes can take based on the analysis of facial data in section. 2.2. The node sizes for the rest of the nodes were provided to be $3$ signifying low, medium and high states for the corresponding nodes.

\paragraph{Learning and Inference} The structure of the $2$ time-slice \ac{DBN} comprises within time-slice connections, that are present between the nodes of the same time-slice and between time-slice connections, that are present between the nodes of the time-slice $t$ and $t+1$. The structure of the \ac{DBN} with the nodes defined in Fig. \ref{fig:Nodes}
was learnt using \acf{mmhc} algorithm, a heuristic algorithm that performs a statistical sieving of the search space followed by a greedy evaluation \cite{tsamardinos2006max}. The method also used BDeu (Bayesian-Dirichlet equivalent uniform) scoring function to search and score the networks. To learn within time-slice structure, prior knowledge was incorporated in the form of layers within the nodes of a time-slice, where the nodes in layer $j$ can have parents only in layers $i \leq j$. The personality scores and the baseline wellbeing score were designated as the first layer as they represent the initial state of the participants, AU states, i.e. $AU\_med$ and $AU\_int$ as the second layer as they represent participants' experience during the sessions and session experience ratings as the third/last layer as they represent outcomes of the interaction as shown in Fig. \ref{fig:Nodes}. Parameter learning was then performed where the parameters of the conditional probability distributions of the nodes given the learnt structure and the data were learnt using maximum-a-posteriori (MAP) estimate of the parameters \cite{murphy2002dynamic}. After learning the DBN, we performed evidence-based forecasting of the session ratings at a new time-step based on the partial observation of personality, baseline wellbeing and the states of $AU\_med$ and $AU\_int$ and the complete observation of the previous time step. We construct an inference engine using the DBN learnt from the first 3 time steps. Expectation-Maximization (EM) algorithm \cite{koller2009probabilistic} was used to update the inference engine and predict values of the unobserved nodes based on the partial observation.

\section{Results}

\subsection{Learnt Structure} 
\begin{figure*}[h]
  \centering
  \includegraphics[width=0.75\textwidth]{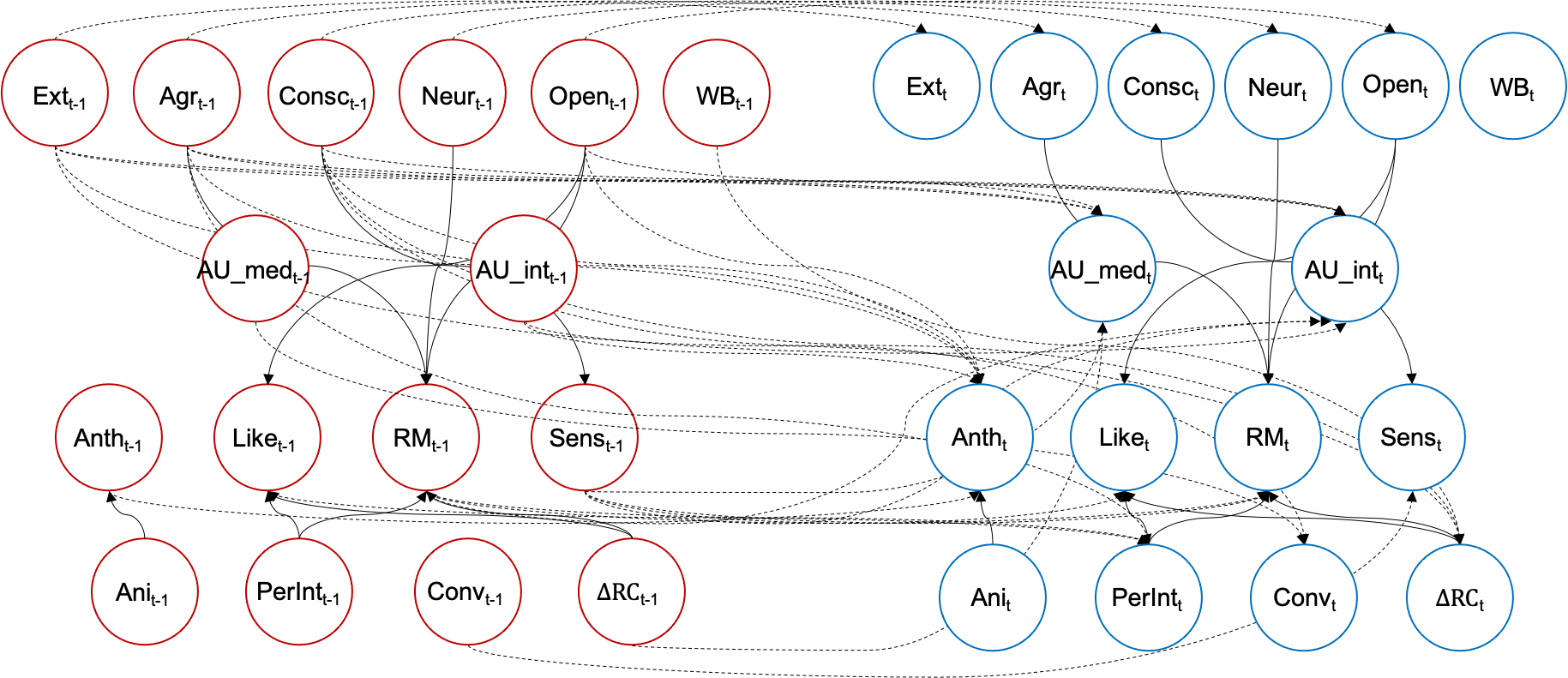}
  \caption{Learnt structure of the $2$ time-slice \ac{DBN} using data from first $3$ time-points i.e. $t1 \rightarrow t2 + t2 \rightarrow t3$, ($+$ denotes concatenation; total $18$ observations) to predict session ratings for $t=4$ and $t=5$. Nodes of $t-1$ are shown in red and those of $t$ are shown in blue. Within time-slice connections are shown as solid lines and between time-slice connections are shown in dotted line.}
  \label{fig:Structure-bnstruct}
\end{figure*}

\begin{figure}
  \centering
  \includegraphics[width=0.9\columnwidth]{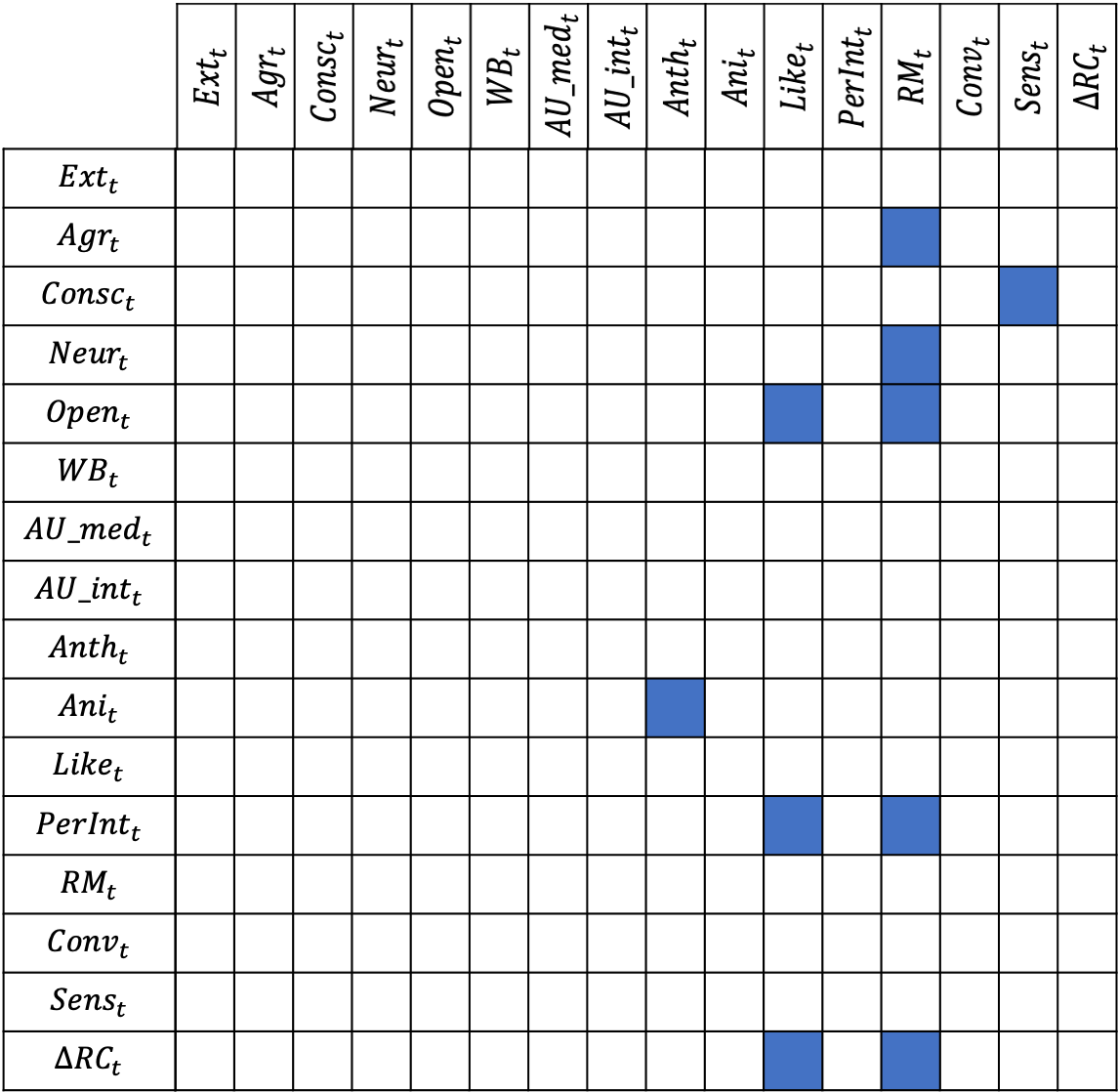}
  \caption{Within time-slice structure.}
  \label{fig:within_adjacency}
  \vspace{-2mm}
\end{figure}

\begin{figure}
  \centering
  \includegraphics[width=0.9\columnwidth]{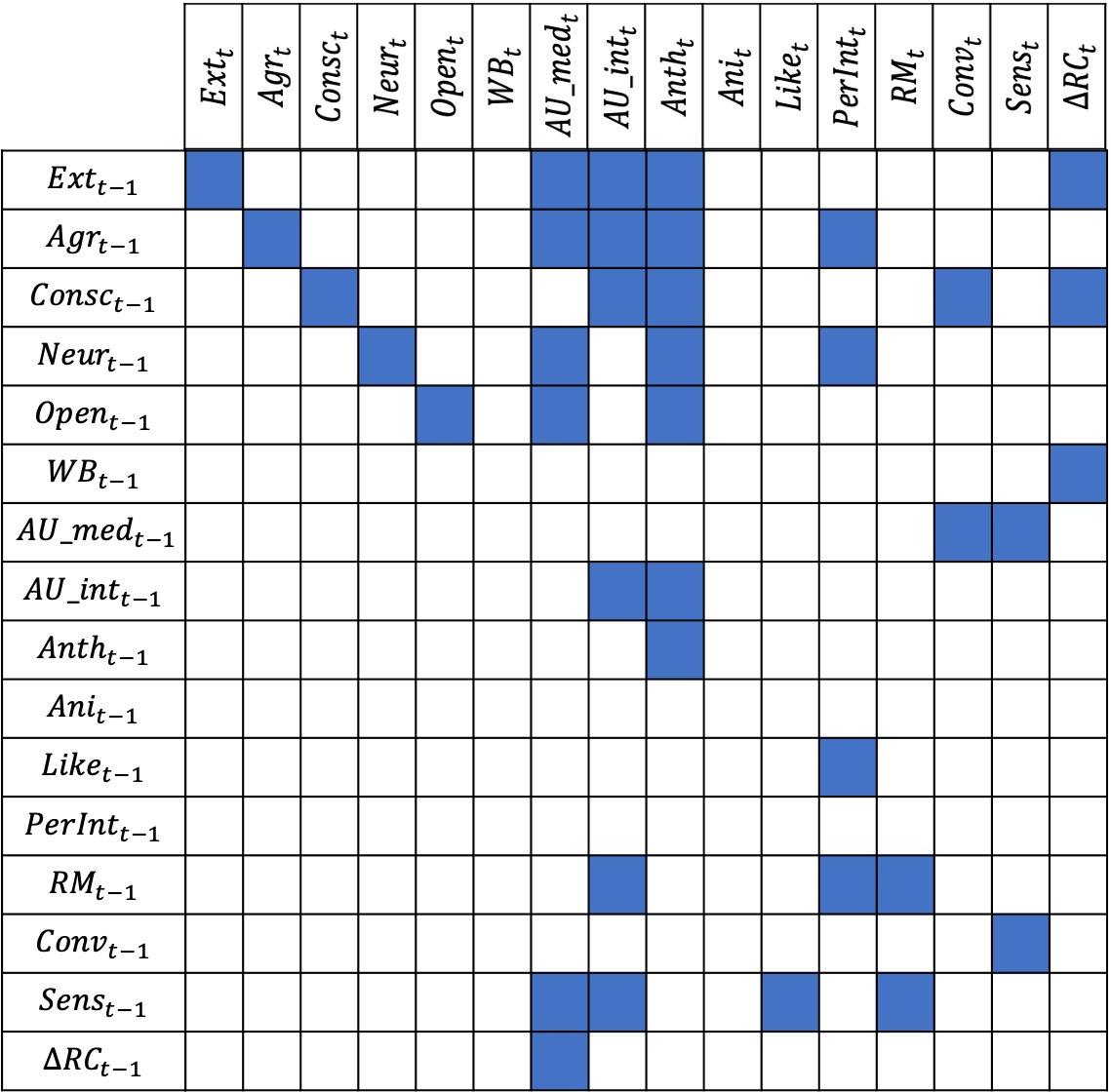}
  \caption{Between time-slice structure.}
  \label{fig:between_adjacency}
  \vspace{-2mm}
\end{figure}

The structure and parameter learning of the 2 time-slice DBN is performed using 2 time-slice data from the first two, and the second and third time points concatenated ($t1 \rightarrow t2 + t2 \rightarrow t3$, $+$ stands for concatenation; total of $18$ observations). This is to ensure that there are sufficient number of data points to execute the inference algorithm. The learnt structure of the 2 time-slice \ac{DBN} is shown in Fig. \ref{fig:Structure-bnstruct}. The adjacency matrices for the within and between time slice structures were also shown in Figs. \ref{fig:within_adjacency} and \ref{fig:between_adjacency} respectively. We discuss noted features of the network below based on the directional edges found in the within and between time-slice structures.

\paragraph{Edges From the Personality Variables}
Multiple edges were found from the personality scores of the participants across the dimensions of \textit{Extroversion, Agreeableness, Conscientiousness, Neuroticism} and \textit{Openness} to the AU states as well as on the session ratings. Within time-slice edges were found from \textit{Agreeableness, Conscientiousness} and \textit{Neuroticism} to the \textit{Robot Motion} within each time-slice suggesting that the ratings for the latter are influenced by participant's personality. \textit{Openness} was found to influence \textit{Likeability} suggesting that the people's willingness to be open towards the idea of a robot coach influenced their ratings about how much they liked the robot. \textit{Conscientiousness} was found to influence \textit{Sensations}, i.e. how much the participants enjoyed the sessions overall. 

Personality dimensions were also found to have longitudinal influence on the $AU\_med$ and $AU\_int$ variables representing the states of facial AUs during \textit{meditation} and \textit{interaction} phases respectively. $AU\_med$ was found to be influenced by \textit{Extroversion, Agreeableness, Neuroticism} and \textit{Openness} while $AU\_int$ was found to be influenced by \textit{Extroversion, Agreeableness}  and \textit{Conscientiousness}. These influences suggest that the emotional experiences of the participants during mindfulness meditations and discussions are also dependent on the personality dimensions, e.g. how open they are about trying mindfulness meditations, whether they exhibit high/low neuroticism and extroversion levels and so on. All personality dimensions also showed between time-slice influences on the \textit{Anthropomorphism} ratings indicating that personality plays a role in attributing anthropomorphic characteristics to the robot coach over time. \textit{Perceived Intelligence} was found to be influenced by \textit{Neuroticism} and \textit{Agreeableness}, i.e. the participants disposition to be suspicious or be convinced about the intelligence levels of the robot coach conducting mindfulness sessions plays an important role in its acceptance. Between time-slice influences were also found from \textit{Conscientiousness} to \textit{Conversation} and $\Delta RC$ ratings. 

\paragraph{Edges From the Facial AU Displays} The facial AU states exhibited during \textit{meditation} and \textit{interaction} phases influenced the session ratings. We found between time-slice connections from $AU\_med$ to \textit{Conversation} and \textit{Sensations} ratings suggesting that how the participants' feel about the sessions and rate the conversations during the sessions evolve based on their AU displays during \textit{meditation} phase of the sessions. We also found between time-slice connection from $AU\_int_{t-1}$ to $AU\_int_t$ suggesting that the facial AU displays during interactions kept evolving across sessions. $AU\_int$ was also found to influence the \textit{Anthropomorphism} ratings across time suggesting that the facial AUs displayed during the interaction with the robot coach also contribute to the participants attributing anthropomorphic characteristics to it.

\paragraph{Edges Involving the Perception Ratings of the Coach} We found edges between the items of the session ratings in both within and between time-slice structures. \textit{Animacy} ratings within a session were found to influence \textit{Anthropomorphism} ratings. \textit{Anthropomorphism} ratings were found to be longitudinally evolving as evidenced by the presence of an edge from $Anth_{t-1}$ to $Anth_t$ nodes. A closed-loop structure was found between \textit{Perceived Intelligence} and \textit{Likeability} ratings (i.e. $PerInt_{t-1} \rightarrow Like_{t-1} \rightarrow PerInt_t$) suggesting that these aspects closely influence each other across time. \textit{Robot Motion} ratings were seen to be evolving longitudinally as seen from the edge from $RM_{t-1}$ to $RM_t$ suggesting that these ratings change over multiple sessions as a function of how well the robot moves. In addition to the personality dimensions, \textit{Robot motion} was found to be influenced by \textit{Perceived Intelligence} ratings within each session and by \textit{Sensation} ratings across sessions. These influences reveal that the ratings for \textit{Robot Motion} also evolve as a result of how the people feel and connect to the session content and the robot coach. \textit{Conversation ratings}, i.e. how well the participants were able to discuss with the robot coach influenced the \textit{Sensation} ratings, i.e. how they felt about the sessions. We also found between time-slice edges from \textit{Sensation} ratings to $AU\_med$, $AU\_int$ suggesting how the participants felt about a session influenced the facial displays during the following sessions.

\paragraph{Edges Involving $\Delta RC$ Ratings}
The $\Delta RC$ ratings signify the effectiveness of the mindfulness sessions in making the participants relaxed and calm after the session compared to before the session. We found that the longitudinal evolution of the $\Delta RC$ ratings were influenced by \textit{Extroversion, Conscientiousness} and baseline wellbeing values obtained using the \textit{WEMWBS} questionnaire. The $\Delta RC$ ratings were in turn found to effect the \textit{Likeability} and the facial AUs during the \textit{meditation} phase in the future sessions suggesting that the $\Delta RC$ ratings are crucial in determining how much participants liked the robot coach and whether they become more engaged in the meditation practices in further sessions.




\subsection{Inference}
EM algorithm was used to predict the session ratings for the remaining two time steps (i.e. $t=4$ and $t=5$) of our 5-week longitudinal data. The predicted values at $t=4$ and $t=5$ are shown in Table. \ref{table:infer}. The values of the unobserved nodes were predicted for each subject and time-point separately. We also provide two subject-wise metrics -- root mean squared errors (RMSE) and R-squared $(R^2)$ score, to understand how good the predicted values are compared to the actual values. $R^2$ score is a standard metric used to measure the goodness-of-fit in regression analysis and is defined as $R^2 = 1- \frac{g}{s}$, where $g = \Sigma((y_{orig}-y_{pred})^2)$ and $s = \Sigma((y_{orig}-mean(y_{orig}))^2)$. It lies typically between 0 and 1 (the closer to 1 the better), and can also be negative on unseen data although this indicates a poor model. The learnt \ac{DBN} performs well on most of the subjects except for both the time points (i.e. $t=4$ and $t=5$) for S3 and for $t=5$ for S6 and S8 as indicated by the negative $R^2$ scores. 

\begin{table*}[ht]
\small
\centering
  \caption{Forecasting future time-steps for each subject (S) at $t=4$ (1st row) and $t=5$ (2nd row) using  learnt \ac{DBN}. Predicted value followed by the actual value shown in parantheses are presented nodes corresponding to the session ratings. The regression metrics, $RMSE$ and $R^2$ score, were provided for each subject to indicate the goodness-of-fit.}
  {
  \begin{tabular}{lllllllllll}
  \Xhline{1.0 pt}
    \makecell[l]{} &
    \makecell[l]{\textbf{$Anth_t$}} & \makecell[l]{\textbf{$Ani_t$}} & \makecell[l]{\textbf{$Like_t$}} & \makecell[l]{\textbf{$PerInt_t$}}&
    \makecell[l]{\textbf{$RM_t$}} & \makecell[l]{\textbf{$Conv_t$}} & \makecell[l]{\textbf{$Sens_t$}} & \makecell[l]{\textbf{$\Delta RC_t$}} &
    \makecell[l]{\textbf{$RMSE$}} &
    \makecell[l]{{$R^2$ score}}\\ \Xhline{1.0 pt}
    \makecell[l]{S1}&
    \makecell[l]{1.83(1.8)\\2.59(2)}&
    \makecell[l]{2.54(1.83)\\2.56(2)}&
    \makecell[l]{3.75(3.4)\\3.04(3.6)}&
    \makecell[l]{3.99(2.2)\\3.37(2.2)}&
    \makecell[l]{2.84(2)\\4.02(2.5)}&
    \makecell[l]{3.66(4.25)\\3.31(4.75)}&
    \makecell[l]{4.85(5)\\4.34(5)}&
    \makecell[l]{0.5(1.5)\\1.12(2)}&
    \makecell[l]{0.86\\0.99}&
    \makecell[l]{0.5024\\0.2955}\\
    \Xhline{0.5 pt}
    \makecell[l]{S2}&
    \makecell[l]{3.55(3.2)\\4.35(2.4)}&
    \makecell[l]{2.53(3)\\2.98(3.17)}&
    \makecell[l]{3.96(4)\\3.9(4.6)}&
    \makecell[l]{3.8(4)\\3.8(4.6)}&
    \makecell[l]{4.37(3)\\4.28(3)}&
    \makecell[l]{4.09(4.25)\\3.47(4.25)}&
    \makecell[l]{4.27(4)\\4.27(4.33)}&
    \makecell[l]{1.67(2)\\0.97(1)}&
    \makecell[l]{0.55\\0.95}&
    \makecell[l]{0.401\\0.367}\\
    \Xhline{0.5 pt}
    \makecell[l]{S3}& 
    \makecell[l]{3.39(4.4)\\3.8(3.4)}&
    \makecell[l]{1.86(4.67)\\2.25(3.17)}&
    \makecell[l]{3.1(5)\\4.05(4)}&
    \makecell[l]{4.32(4.6)\\2.39(4.8)}&
    \makecell[l]{1.66(4)\\1.56(4.5)}&
    \makecell[l]{4.69(5)\\4.94(4.75)}&
    \makecell[l]{4.73(5)\\4.29(5)}&
    \makecell[l]{1.36(1.5)\\0.33(0.5)}&
    \makecell[l]{1.51\\1.42}&
    \makecell[l]{-0.8986\\-0.0489}\\
    \Xhline{0.5 pt}
    \makecell[l]{S4}& 
    \makecell[l]{2.76(3.2)\\2.79(3.4)}&
    \makecell[l]{2.24(3.67)\\1.87(4)}&
    \makecell[l]{4.82(4)\\4.13(4)}&
    \makecell[l]{4.48(3)\\3.47(3)}&
    \makecell[l]{1.97(4)\\3.21(4.5)}&
    \makecell[l]{4.1(5)\\4.49(5)}&
    \makecell[l]{3.78(5)\\4.81(5)}&
    \makecell[l]{1.62(0)\\1.9(2)}&
    \makecell[l]{1.31\\0.94}&
    \makecell[l]{0.2238\\0.0508}\\
    \Xhline{0.5 pt}
    \makecell[l]{S5}& 
    \makecell[l]{1.87(1.4)\\2.44(1.6)}&
    \makecell[l]{2.64(2)\\2.6(2)}&
    \makecell[l]{4.26(4)\\3.6(3.8)}&
    \makecell[l]{3.99(4.8)\\4.94(5)}&
    \makecell[l]{2.03(2.5)\\3.53(1.5)}&
    \makecell[l]{4.3(4)\\4.35(4.25)}&
    \makecell[l]{4.14(3.75)\\3.45(4)}&
    \makecell[l]{0.6(1)\\0.78(2)}&
    \makecell[l]{0.5\\0.94}&
    \makecell[l]{0.8538\\0.473}\\
    \Xhline{0.5 pt}
    \makecell[l]{S6}& 
    \makecell[l]{3.48(3)\\3.86(4.6)}&
    \makecell[l]{1.92(2.67)\\2.48(4.4)}&
    \makecell[l]{4.12(4.2)\\3.9(5)}&
    \makecell[l]{2.53(5)\\3.12(5)}&
    \makecell[l]{2.81(3)\\2.32(4.5)}&
    \makecell[l]{4.14(5)\\4.43(4.75)}&
    \makecell[l]{4.03(5)\\4.77(5)}&
    \makecell[l]{0.86(0)\\1.77(2)}&
    \makecell[l]{1.08\\1.32}&
    \makecell[l]{0.5495\\-0.983}\\
    \Xhline{0.5 pt}
    \makecell[l]{S7}& 
    \makecell[l]{2.91(2.4)\\2.86(3.4)}&
    \makecell[l]{2.12(2.5)\\1.87(3.2)}&
    \makecell[l]{4.58(3.2)\\3.55(4)}&
    \makecell[l]{4.95(3)\\3.21(4)}&
    \makecell[l]{3.46(2.5)\\1.89(4)}&
    \makecell[l]{4.57(4.75)\\4.48(5)}&
    \makecell[l]{4.6(5)\\4.44(5)}&
    \makecell[l]{1.56(1)\\0.11(1)}&
    \makecell[l]{0.97\\1.042}&
    \makecell[l]{0.369\\0.2293}\\
    \Xhline{0.5 pt}
    \makecell[l]{S8}& 
    \makecell[l]{3.7(2.6)\\4.25(3.8)}&
    \makecell[l]{1.95(2.5)\\2.13(3.8)}&
    \makecell[l]{4.22(3.4)\\4.1(4)}&
    \makecell[l]{3.73(3.4)\\4.72(3)}&
    \makecell[l]{2.5(3)\\1.92(3)}&
    \makecell[l]{4.69(5)\\3.72(4)}&
    \makecell[l]{3.78(3.75)\\4.07(4)}&
    \makecell[l]{0.69(1)\\0.01(1)}&
    \makecell[l]{0.59\\1.01}&
    \makecell[l]{0.702\\-0.0989}\\
    \Xhline{0.5 pt}
    \makecell[l]{S9}& 
    \makecell[l]{3.64(2.6)\\4.3}&
    \makecell[l]{1.89(3.17)\\2.88}&
    \makecell[l]{3.09(4)\\3.92}&
    \makecell[l]{3.06(4)\\4.82}&
    \makecell[l]{2.33(1.5)\\3.98}&
    \makecell[l]{4.46(3.5)\\4.79}&
    \makecell[l]{3.34(3.75)\\3.94}&
    \makecell[l]{0.7(0.5)\\1.56}&
    \makecell[l]{0.88\\--}&
    \makecell[l]{0.4504\\--}\\
    \Xhline{1.0 pt}
  \end{tabular}
  }
\label{table:infer}  

\end{table*}

\subsection{Applications of the Learnt DBN}
In this section we provide two examples where the learnt DBN can be applied. Firstly, the learnt probabilistic model is useful in meaningfully imputing the missing values in the dataset. For example, participant S9 did not attend the mindfulness session for week 5. However, as seen in Table. \ref{table:infer}, the second row of S9 presents an estimation of the values of session ratings. This is very helpful in the case of longitudinal studies where participants often miss one or more sessions. Secondly, the learnt DBN is useful in predicting longitudinal trends for new participants given initial conditions such as their personality scores. As demonstrated in Table. \ref{tab:new_pers} (a), we define two new personality profiles, $C1$ and $C2$ with low and high conscientiousness scores respectively. Using the learnt DBN, we estimate the ratings for all $5$ time steps of $C1$ and $C2$ (Table. \ref{tab:new_pers} (b)). The estimated longitudinal ratings were found to reflect the learnt structure in Sec. 4.1. For example, estimated \textit{Sensations} ratings ($mean = 4.634$) for $C2$ with high Conscientiousness were found to be  higher than the estimated  \textit{Sensations} ratings ($mean = 3.567$) for $C1$ with low Conscientiousness. This is in line with the observation that \textit{Conscientiousness} was found to have within session influence on \textit{Sensations} ratings. Further, there is a steady increase in the \textit{Conversation} ratings and a steady decrease in the $\Delta RC$ ratings for $C2$ compared to $C1$ which is also supported by the longitudinal influences found from \textit{Conscientiousness} to \textit{Conversation} and $\Delta RC$ ratings. Thus, the DBN acts as a reliable model to infer longitudinal behaviour of new participants. 

\begin{table*}[h]
\centering
\caption{Demonstrating longitudinal inference of session ratings of a new participant given their personality profile. In (a), we present examples of  personality profiles of two new participants, $C1$ and $C2$ with low ($Consc=1$) and high ($Consc=5$) \textit{Conscientiousness} values respectively. The rest of the personality dimensions were set to mid-value $3$ on the scale, $1-5$. In (b), we present the estimated longitudinal session ratings of $C1$ and $C2$.}
        \centering
        \subcaption{Personality profiles with low ($C1$) and high ($C2$) Conscientiousness values.}
        \vspace{2mm}
        \begin{tabular}{cccccc}
        \Xhline{1.0 pt}
        {} & $Ext$ & $Agr$ & $Consc$ & $Neur$ & $Open$ \\
        \Xhline{1.0 pt}
        $C1$ (low $Consc$) & $3$ & $3$ & $1$ & $3$ & $3$ \\
        \Xhline{0.5 pt}
        $C2$ (high $Consc$) & $3$ & $3$ & $5$ & $3$ & $3$ \\
        \Xhline{1.0 pt}
       \end{tabular}
        \vspace{5mm}

        \centering
        \subcaption{Estimated longitudinal session ratings of $C1$ and $C2$ for all 5 sessions using the learnt DBN.}
        \vspace{2mm}
        \begin{tabular}{llllllllll}
        \Xhline{1.0 pt}
        {} & $Time$ & $Anth$ & $Ani$ & $Like$ &
        $PerInt$ & $RM$ & $Conv$ & $Sens$ & $\Delta RC$\\
        \Xhline{1.0 pt}
        $C1$ & $t=1$ & $2.568$ & $3.381$ & $3.750$ & $2.236$ & $1.565$ & $3.870$ & $3.800$ & $0.292$\\
        \Xhline{0.5 pt}
        $C1$ & $t=2$ & $2.758$ & $2.995$ & $3.102$ & $3.720$ & $2.077$ & $4.668$ & $3.386$ & $0.067$\\
        \Xhline{0.5 pt}
        $C1$ & $t=3$ & $4.632$ & $2.981$ & $3.704$ & $4.280$ & $1.570$ & $4.562$ & $3.214$ & $0.496$\\
        \Xhline{0.5 pt}
        $C1$ & $t=4$ & $3.111$ & $1.878$ & $3.116$ & $4.922$ & $1.807$ & $4.930$ & $4.113$ & $0.232$\\
        \Xhline{0.5 pt}
        $C1$ & $t=5$ & $2.820$ & $2.765$ & $3.283$ & $4.632$ & $2.151$ & $4.735$ & $3.318$ & $0.783$\\
        \Xhline{0.5 pt}
        $C2$ & $t=1$ & $2.405$ & $4.042$ & $3.865$ & $3.327$ & $1.046$ & $4.175$ & $3.488$ & $0.734$\\
        \Xhline{0.5 pt}
        $C2$ & $t=2$ & $2.030$ & $2.409$ & $3.184$ & $3.448$ & $1.628$ & $4.558$ & $4.981$ & $0.614$\\
        \Xhline{0.5 pt}
        $C2$ & $t=3$ & $2.874$ & $2.321$ & $3.246$ & $4.833$ & $2.077$ & $4.879$ & $4.976$ & $0.305$\\
        \Xhline{0.5 pt}
        $C2$ & $t=4$ & $2.742$ & $2.736$ & $3.212$ & $4.164$ & $2.017$ & $4.895$ & $4.881$ & $0.816$\\
        \Xhline{0.5 pt}
        $C2$ & $t=5$ & $2.740$ & $2.422$ & $3.020$ & $4.328$ & $1.675$ & $4.971$ & $4.844$ & $0.122$\\
        \Xhline{1.0 pt}
        \end{tabular}
     
     \label{tab:new_pers}

\end{table*}

\section{Discussion and Conclusion}
\subsection{Discussion}
Our learning-based approach using \ac{DBN}, provided an explanatory model that helped us understand multiple facets of human-robot interactions and how they change longitudinally during robot-delivered mindfulness sessions. The model provides an account of how participant-related factors such as personality, initial wellbeing and facial AU states elicited during the sessions influence their session ratings, i.e. perception ratings of the coach and the change in their self-reported relaxation/calm levels. The learnt structure demonstrates the longitudinal evolution of variables such as $AU\_int$, \textit{Anthropomorphism} and \textit{Robot Motion} suggesting the importance of longitudinal studies in completely realizing how participants feel about various aspects of the robot design. Each personality dimension was shown to have multiple influences on the displayed facial AUs during sessions, the perception ratings of the robot coach and the changes in relaxation/calm levels before and after the sessions. These findings are also supported by our previous study \cite{bodala2021teleoperated}, where we found that \textit{Robot Motion} ratings improves longitudinally over time and the personality dimensions, \textit{Conscientiousness} and \textit{Neuroticicsm} affect the ratings on \textit{Robot Motion} and \textit{Sensations}. This suggests that it is crucial to consider user characteristics such as their personality in designing technological interventions for wellbeing to ensure acceptance and adherence in long-term. Participants characteristics such as \textit{Openness} and \textit{Agreeableness} were also found to be key in influencing the ratings on \textit{Likeability, Perceived Intelligence} and \textit{Robot Motion}. This is also supported by \cite{axelsson2021participatory}, where discussions with participants revealed that openness towards evidence-based change influenced participants attitudes towards a robot wellbeing coach.

The model also provided insights into within session and long-term influences between various aspects of the perceptions ratings of the coach. For example, \textit{Robot Motion} was found to be influenced by \textit{Perceived Intelligence} and $\Delta RC$ ratings within each session but was influenced by past \textit{Robot Motion} and \textit{Sensations} ratings across sessions. This suggests that the instantaneous influences and long-ranging influences on a variable may vary and might provide important insights for the roboticists in terms of design priorities. For example, it may be important to prioritize design in terms of functionality to appear intelligent, but it is equally important to ensure that the participants enjoy the interactions in addition to the robot's functionality. This is further demonstrated by the closed loop influences \textit{Perceived Intelligence} and \textit{Likeability} exert on each other across time. 
\vspace{-3mm}
\subsection{Limitations and Future Work}
In our paper, the specific network structure of the model is learnt based on the available self-reported scores and facial AU states which summarize each weekly interaction in one time-slice. Further, it is not evident which emotional states correspond to the facial AU states displayed during the meditation and interaction sub-sessions, i.e the values of the $AU\_med$ and $AU\_int$ nodes. Future work could extend this learning-based approach to jointly learn more fine-grained associations and investigate further on the emotional states corresponding to the facial AU states displayed. However, this investigation will need further annotations of the dataset at finer time-scales. Moreover, fine-grained structures imply more nodes in the graphical model and need for more participants' data to achieve better representation of the interactions.

\subsection{Conclusion}
In this paper, we present a longitudinal graphical model that serves as an intuitive way of modelling complex, longitudinal relationships between multiple variables of interest. The 2 time-slice DBN is trained using $3$ weeks and is used to predict the session ratings of the remaining $2$ weeks in our 5-weeks longitudinal study. Further, the two applications of the learnt DBN, i.e., imputation of missing data and estimating longitudinal trends in session ratings of a new participant given their personality profile are demonstrated. Both these applications of the learnt DBN can help in synthesizing new large datasets without missing values that can be useful in reliable analysis and understanding of HRI scenarios without collecting large-scale longitudinal data which is expensive and time-consuming. Further, this model can act as a good foundation to implement further experiments and can be updated as new data becomes available.


 
\section*{Funding}
This  work  is  supported  by  the Engineering and Physical Sciences Research Council (EPSRC)  under  grant  reference  EP/R030782/1.



\begin{acronym}
    \acro{SAR}{Socially Assistive Robot}
    \acro{HRI}{human-robot interaction}
    \acro{DBN}{Dynamic Bayesian Network}
    \acro{BN}{Bayesian Network}
    \acro{WEMWBS}{Warwick-Edinburgh Mental Wellbeing
    Scale}
    \acro{HC}{Human Coach}
    \acro{RC}{teleoperated Robot Coach}
    \acro{AUs}{facial action units}
    \acro{mmhc}{max-min hill climbing}
    \acro{HRI}{human-robot interaction}
    \acro{GLM}{generalized linear model}
\end{acronym}

\bibliographystyle{IEEEtran}
\bibliography{main}

\begin{thebibliography}{10}
\providecommand{\url}[1]{#1}
\csname url@samestyle\endcsname
\providecommand{\newblock}{\relax}
\providecommand{\bibinfo}[2]{#2}
\providecommand{\BIBentrySTDinterwordspacing}{\spaceskip=0pt\relax}
\providecommand{\BIBentryALTinterwordstretchfactor}{4}
\providecommand{\BIBentryALTinterwordspacing}{\spaceskip=\fontdimen2\font plus
\BIBentryALTinterwordstretchfactor\fontdimen3\font minus
  \fontdimen4\font\relax}
\providecommand{\BIBforeignlanguage}[2]{{%
\expandafter\ifx\csname l@#1\endcsname\relax
\typeout{** WARNING: IEEEtran.bst: No hyphenation pattern has been}%
\typeout{** loaded for the language `#1'. Using the pattern for}%
\typeout{** the default language instead.}%
\else
\language=\csname l@#1\endcsname
\fi
#2}}
\providecommand{\BIBdecl}{\relax}
\BIBdecl

\bibitem{feil2005defining}
D.~Feil-Seifer and M.~J. Mataric, ``Defining socially assistive robotics,'' in
  \emph{9th International Conference on Rehabilitation Robotics, 2005. ICORR
  2005.}\hskip 1em plus 0.5em minus 0.4em\relax IEEE, 2005, pp. 465--468.

\bibitem{conti2021brief}
D.~Conti, S.~Di~Nuovo, and A.~Di~Nuovo, ``A brief review of robotics
  technologies to support social interventions for older users,'' \emph{Human
  Centred Intelligent Systems}, pp. 221--232, 2021.

\bibitem{lima2021robotic}
M.~R. Lima, M.~Wairagkar, N.~Natarajan, S.~Vaitheswaran, and R.~Vaidyanathan,
  ``Robotic telemedicine for mental health: a multimodal approach to improve
  human-robot engagement,'' \emph{Frontiers in Robotics and AI}, vol.~8, 2021.

\bibitem{jeong2020robotic}
S.~Jeong, S.~Alghowinem, L.~Aymerich-Franch, K.~Arias, A.~Lapedriza, R.~Picard,
  H.~W. Park, and C.~Breazeal, ``A robotic positive psychology coach to improve
  college students’ wellbeing,'' in \emph{IEEE RO-MAN}, 2020, pp. 187--194.

\bibitem{bodala2021teleoperated}
I.~P. Bodala, N.~Churamani, and H.~Gunes, ``Teleoperated robot coaching for
  mindfulness training: A longitudinal study,'' in \emph{2021 30th IEEE
  International Conference on Robot \& Human Interactive Communication
  (RO-MAN)}.\hskip 1em plus 0.5em minus 0.4em\relax IEEE, 2021, pp. 939--944.

\bibitem{gockley2005designing}
R.~Gockley, A.~Bruce, J.~Forlizzi, M.~Michalowski, A.~Mundell, S.~Rosenthal,
  B.~Sellner, R.~Simmons, K.~Snipes, A.~C. Schultz \emph{et~al.}, ``Designing
  robots for long-term social interaction,'' in \emph{IEEE/RSJ IROS}.\hskip 1em
  plus 0.5em minus 0.4em\relax IEEE, 2005, pp. 1338--1343.

\bibitem{de2017they}
M.~De~Graaf, S.~B. Allouch, and J.~Van~Diik, ``Why do they refuse to use my
  robot?: Reasons for non-use derived from a long-term home study,'' in
  \emph{2017 12th ACM/IEEE International Conference on Human-Robot Interaction
  (HRI}.\hskip 1em plus 0.5em minus 0.4em\relax IEEE, 2017, pp. 224--233.

\bibitem{paetzel2020persistence}
M.~Paetzel, G.~Perugia, and G.~Castellano, ``The persistence of first
  impressions: The effect of repeated interactions on the perception of a
  social robot,'' in \emph{Proceedings of the 2020 ACM/IEEE International
  Conference on Human-Robot Interaction}, 2020, pp. 73--82.

\bibitem{winkle2018social}
K.~Winkle, P.~Caleb-Solly, A.~Turton, and P.~Bremner, ``Social robots for
  engagement in rehabilitative therapies: Design implications from a study with
  therapists,'' in \emph{Proc. ACM/IEEE HRI}, 2018, pp. 289--297.

\bibitem{de2020towards}
E.~J. De~Visser, M.~M. Peeters, M.~F. Jung, S.~Kohn, T.~H. Shaw, R.~Pak, and
  M.~A. Neerincx, ``Towards a theory of longitudinal trust calibration in
  human--robot teams,'' \emph{International journal of social robotics},
  vol.~12, no.~2, pp. 459--478, 2020.

\bibitem{cespedes2021socially}
N.~C{\'e}spedes~G{\'o}mez, B.~Irfan, E.~Senft, C.~A. Cifuentes, L.~F.
  Gutierrez, M.~Rincon-Roncancio, T.~Belpaeme, and M.~Munera, ``A socially
  assistive robot for long-term cardiac rehabilitation in the real world,''
  \emph{Frontiers in Neurorobotics}, vol.~15, p.~21, 2021.

\bibitem{scutari2009learning}
M.~Scutari, ``Learning bayesian networks with the bnlearn r package,''
  \emph{arXiv preprint arXiv:0908.3817}, 2009.

\bibitem{murphy2002dynamic}
K.~P. Murphy, \emph{Dynamic bayesian networks: representation, inference and
  learning}.\hskip 1em plus 0.5em minus 0.4em\relax University of California,
  Berkeley, 2002.

\bibitem{huang2014learning}
C.-M. Huang and B.~Mutlu, ``Learning-based modeling of multimodal behaviors for
  humanlike robots,'' in \emph{2014 9th ACM/IEEE International Conference on
  Human-Robot Interaction (HRI)}.\hskip 1em plus 0.5em minus 0.4em\relax IEEE,
  2014, pp. 57--64.

\bibitem{leite2013social}
I.~Leite, C.~Martinho, and A.~Paiva, ``Social robots for long-term interaction:
  a survey,'' \emph{International Journal of Social Robotics}, vol.~5, no.~2,
  pp. 291--308, 2013.

\bibitem{perugia2021can}
G.~Perugia, M.~Paetzel-Pr{\"u}smann, M.~Alanenp{\"a}{\"a}, and G.~Castellano,
  ``I can see it in your eyes: Gaze as an implicit cue of uncanniness and task
  performance in repeated interactions with robots,'' \emph{Frontiers in
  Robotics and AI}, vol.~8, 2021.

\bibitem{paetzel2019let}
M.~Paetzel and G.~Castellano, ``Let me get to know you better: can interactions
  help to overcome uncanny feelings?'' in \emph{Proceedings of the 7th
  International Conference on Human-agent Interaction}, 2019, pp. 59--67.

\bibitem{ployhart2011quick}
R.~E. Ployhart and A.-K. Ward, ``The “quick start guide” for conducting and
  publishing longitudinal research,'' \emph{Journal of Business and
  Psychology}, vol.~26, no.~4, pp. 413--422, 2011.

\bibitem{spangler2020multilevel}
D.~Spangler, S.~Alam, S.~Rahman, J.~Crone, R.~Robucci, N.~Banerjee, S.~Kerick,
  and J.~Brooks, ``Multilevel longitudinal analysis of shooting performance as
  a function of stress and cardiovascular responses,'' \emph{IEEE Transactions
  on Affective Computing}, 2020.

\bibitem{mcduff2019longitudinal}
D.~McDuff, E.~Jun, K.~Rowan, and M.~Czerwinski, ``Longitudinal observational
  evidence of the impact of emotion regulation strategies on affective
  expression,'' \emph{IEEE Transactions on Affective Computing}, 2019.

\bibitem{ong2019applying}
D.~Ong, H.~Soh, J.~Zaki, and N.~Goodman, ``Applying probabilistic programming
  to affective computing,'' \emph{IEEE Transactions on Affective Computing},
  2019.

\bibitem{ong2015affective}
D.~C. Ong, J.~Zaki, and N.~D. Goodman, ``Affective cognition: Exploring lay
  theories of emotion,'' \emph{Cognition}, vol. 143, pp. 141--162, 2015.

\bibitem{wu2018rational}
Y.~Wu, C.~L. Baker, J.~B. Tenenbaum, and L.~E. Schulz, ``Rational inference of
  beliefs and desires from emotional expressions,'' \emph{Cognitive science},
  vol.~42, no.~3, pp. 850--884, 2018.

\bibitem{otsuka2007automatic}
K.~Otsuka, H.~Sawada, and J.~Yamato, ``Automatic inference of cross-modal
  nonverbal interactions in multiparty conversations: " who responds to whom,
  when, and how?" from gaze, head gestures, and utterances,'' in
  \emph{Proceedings of the 9th international conference on Multimodal
  interfaces}, 2007, pp. 255--262.

\bibitem{mihoub2016graphical}
A.~Mihoub, G.~Bailly, C.~Wolf, and F.~Elisei, ``Graphical models for social
  behavior modeling in face-to face interaction,'' \emph{Pattern Recognition
  Letters}, vol.~74, pp. 82--89, 2016.

\bibitem{hong2007mixed}
J.-H. Hong, Y.-S. Song, and S.-B. Cho, ``Mixed-initiative human--robot
  interaction using hierarchical bayesian networks,'' \emph{IEEE Transactions
  on Systems, Man, and Cybernetics-Part A: Systems and Humans}, vol.~37, no.~6,
  pp. 1158--1164, 2007.

\bibitem{montesano2008learning}
L.~Montesano, M.~Lopes, A.~Bernardino, and J.~Santos-Victor, ``Learning object
  affordances: from sensory--motor coordination to imitation,'' \emph{IEEE
  Transactions on Robotics}, vol.~24, no.~1, pp. 15--26, 2008.

\bibitem{song2015task}
D.~Song, C.~H. Ek, K.~Huebner, and D.~Kragic, ``Task-based robot grasp planning
  using probabilistic inference,'' \emph{IEEE transactions on robotics},
  vol.~31, no.~3, pp. 546--561, 2015.

\bibitem{kroenke2001phq}
K.~Kroenke, R.~L. Spitzer, and J.~B. Williams, ``The {PHQ}-9: validity of a
  brief depression severity measure,'' \emph{J. of general internal medicine},
  pp. 606--613, 2001.

\bibitem{spitzer2006brief}
R.~L. Spitzer, K.~Kroenke, J.~B. Williams, and B.~L{\"o}we, ``A brief measure
  for assessing generalized anxiety disorder: The {GAD}-7,'' \emph{Arch. of
  Int. Med.}, pp. 1092--1097, 2006.

\bibitem{galante2018mindfulness}
J.~Galante, G.~Dufour, M.~Vainre, A.~P. Wagner, J.~Stochl, A.~Benton,
  N.~Lathia, E.~Howarth, and P.~B. Jones, ``A mindfulness-based intervention to
  increase resilience to stress in university students (the mindful student
  study): a pragmatic randomised controlled trial,'' \emph{The Lancet Public
  Health}, vol.~3, no.~2, pp. e72--e81, 2018.

\bibitem{topolewska2014short}
E.~Topolewska-Siedzik, E.~Skimina, W.~Strus, J.~Cieciuch, and T.~Rowiński,
  ``The short ipip-bfm-20 questionnaire for measuring the big five,''
  \emph{Ann. of Psych.}, pp. 385--402, 2014.

\bibitem{tennant2007warwick}
R.~Tennant, L.~Hiller, R.~Fishwick, S.~Platt, S.~Joseph, S.~Weich,
  J.~Parkinson, J.~Secker, and S.~Stewart-Brown, ``The warwick-edinburgh mental
  well-being scale (wemwbs): development and uk validation,'' \emph{Health and
  Quality of life Outcomes}, vol.~5, no.~1, p.~63, 2007.

\bibitem{bartneck2009measurement}
C.~Bartneck, D.~Kuli{\'c}, E.~Croft, and S.~Zoghbi, ``Measurement instruments
  for the anthropomorphism, animacy, likeability, perceived intelligence, and
  perceived safety of robots,'' \emph{Int'l J. of Social Robotics}, pp. 71--81,
  2009.

\bibitem{romero2015testing}
A.~Romero-Garc{\'e}s, L.~V. Calderita, J.~Mart{\'\i}nez-G{\'o}mez, J.~P.
  Bandera, R.~Marfil, L.~J. Manso, A.~Bandera, and P.~Bustos, ``Testing a fully
  autonomous robotic salesman in real scenarios,'' in \emph{IEEE ICARSC}, 2015,
  pp. 124--130.

\bibitem{baltruvsaitis2016openface}
T.~Baltru{\v{s}}aitis, P.~Robinson, and L.-P. Morency, ``Openface: an open
  source facial behavior analysis toolkit,'' in \emph{2016 IEEE Winter
  Conference on Applications of Computer Vision (WACV)}.\hskip 1em plus 0.5em
  minus 0.4em\relax IEEE, 2016, pp. 1--10.

\bibitem{franzin2017bnstruct}
A.~Franzin, F.~Sambo, and B.~Di~Camillo, ``bnstruct: an r package for bayesian
  network structure learning in the presence of missing data,''
  \emph{Bioinformatics}, vol.~33, no.~8, pp. 1250--1252, 2017.

\bibitem{tsamardinos2006max}
I.~Tsamardinos, L.~E. Brown, and C.~F. Aliferis, ``The max-min hill-climbing
  bayesian network structure learning algorithm,'' \emph{Machine learning},
  vol.~65, no.~1, pp. 31--78, 2006.

\bibitem{koller2009probabilistic}
D.~Koller and N.~Friedman, \emph{Probabilistic graphical models: principles and
  techniques}.\hskip 1em plus 0.5em minus 0.4em\relax MIT press, 2009.

\bibitem{axelsson2021participatory}
M.~Axelsson, I.~P. Bodala, and H.~Gunes, ``Participatory design of a robotic
  mental well-being coach,'' in \emph{2021 30th IEEE International Conference
  on Robot \& Human Interactive Communication (RO-MAN)}.\hskip 1em plus 0.5em
  minus 0.4em\relax IEEE, 2021, pp. 1081--1088.

\end{thebibliography}


 




\vfill

\end{document}